\documentclass[11pt]{article}

\usepackage[a4paper,margin=1in]{geometry}

\usepackage{amsmath,amssymb}

\usepackage{graphicx}
\usepackage{booktabs}

\usepackage[numbers,sort&compress]{natbib}

\usepackage[colorlinks=true,
            linkcolor=blue,
            citecolor=blue,
            urlcolor=blue]{hyperref}

\usepackage{setspace}
\usepackage{authblk}

\newenvironment{summary}[1][Summary Points]
{\subsection*{#1}}
{}

\title{\bfseries
Biophysics of the Pyrenoid
}

\author[1]{Charley Schaefer}
\author[2]{Mark C. Leake}

\affil[1]{School of Physics and Astronomy \& Astbury Centre for Structural Molecular Biology,
University of Leeds, Leeds LS2 9JT, UK}
\affil[2]{School of Physics, Engineering and Technology,
University of York, York YO10 5DD, UK}

\date{}

\begin{document}

\maketitle
\begin{center}
\small
Preprint under review.\\
Version dated \today.
\end{center}

\begin{center}
\texttt{c.schaefer@leeds.ac.uk}\\
\texttt{mark.leake@york.ac.uk}
\end{center}

\vspace{1em}

\begin{abstract}
Phase-separated liquid droplets organize molecules in cells, but the underlying physical principles differ from abiotic mixing and quantitative rules in living systems remain poorly understood. 
The pyrenoid---a liquid-like organelle that enhances photosynthetic carbon fixation in algae and hornworts---provides an unusually tractable model system. Here, we review recent advances in our understanding of pyrenoids from the perspective of biophysics. We highlight how reaction--diffusion models connect compartment architecture to catalytic performance, how soft matter theories link molecular interactions to condensate assembly, and how modern experimental methods enable these predictions to be tested quantitatively. Recent studies suggest that pyrenoid function may be described by a small number of effective transport and reaction processes, while condensate assembly can be understood through molecular design parameters and thermodynamic constraints. Together, these findings establish the pyrenoid as a powerful system for investigating catalytic compartmentalization, biomolecular self-organization and the emergence of effective physical descriptions in living systems.
\end{abstract}

\newpage

\tableofcontents

\newpage


\section{Introduction}

Living systems frequently control biochemical processes through spatial and temporal organization of cellular material \cite{banani_biomolecular_2017}.
Specifically, the regulation of multiple cellular processes under normal physiological and pathological conditions is often regulated by physicochemical microenvironments composed of spatiotemporal compartments  \cite{roden_rna_2021,alberti_biomolecular_2021, holehouse_molecular_2025}. 
Open biophysics challenges in this field revolve around two questions: how can we relate a compartment's structure to its function, and how is this structure regulated by the cell?
The development of transferable knowledge is largely challenged by the out-of-equilibrium and multicomponent nature of living systems.
Over the past decade, an exemplar biological compartment system of the algal `pyrenoid' has emerged as offering new insights into these questions. 

\begin{figure}[h]
\centering
\includegraphics[width=10cm]{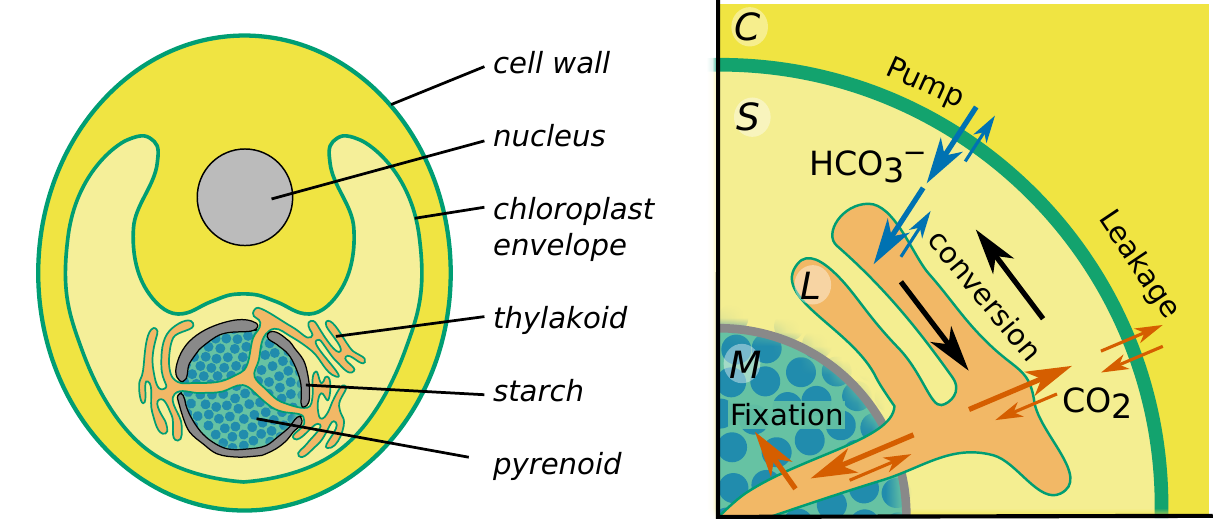}
\caption{Left: schematic representation of the organelles related to the pyrenoid in the green algae C. rheinhardtii. Right: carbon transport from and to the pyrenoid; the letters indicate: C=cytosol; S=stroma; L=lumen; M=matrix.}
\label{fig:Pyrenoid}
\end{figure}

Found in many algae and several hornwort species, pyrenoids are a type of biophysical carbon concentrating mechanism (CCM), where they enhance photosynthetic carbon fixation by co-localizing Ribulose-1,5-bisphosphate carboxylase/oxygenase (Rubisco) with elevated concentrations of inorganic carbon \cite{Villarreal12,barrett_pyrenoids_2021,HeReview23,freeman_rosenzweig_eukaryotic_2017}. 
Their physiological role is therefore comparatively well defined towards enhancing the efficiency of carbon fixation. 
Despite pyrenoids exhibiting remarkable evolutionary and structural diversity, with multi-compartment architectures that vary
substantially between species \cite{barrett_pyrenoids_2021,Catherall2025}, there is a common functional focus driven by consensus biophysics. All pyrenoids 
serve to increase the concentration of CO$_2$ near Rubisco while
reducing losses through diffusion and competing reactions. The repeated
emergence of pyrenoid-like structures therefore suggests that
compartmentalization represents a robust physical solution to the
problem of efficient carbon fixation. Rather than viewing pyrenoids as
a single organelle with a fixed architecture such as depicted in Figure \ref{fig:Pyrenoid}, it may therefore be more
useful to regard them as a family of evolutionary solutions that
implement common transport--reaction principles.
Thus, unlike many condensates whose physiological functions remain debated, the pyrenoid performs a well-defined catalytic role whose output can be quantified directly through photosynthetic carbon fixation, making it an attractive candidate for studying general biophysics principles of both condensates and, more broadly,  cellular organisation.

The pyrenoid was first identified through classical optical microscopy as a refractile object within algal chloroplasts, in combination with classical cell biology, genetics, physiology and biochemical approaches.  Transmission electron microscopy (TEM) subsequently transformed our understanding of its ultrastructure. Thin-section TEM revealed a morphological diversity across algal lineages, from simple spherical condensates to highly elaborate proteinaceous structures integrated with tubular thylakoid membrane systems surrounded by starch plates or other peripheral structures depending on species (see Figure~\ref{fig:Pyrenoid}). Immunogold TEM subsequently established that Rubisco is the principal structural constituent of the pyrenoid matrix \cite{LacosteRoyal1987} and that its organization changes under CCM-inducing conditions \cite{Borkhsenious1998}, demonstrating that pyrenoids are dynamic rather than static structures.
TEM has also been widely used to localize pyrenoid-associated proteins at subcompartment resolution. Carbonic anhydrases, Rubisco activase proteins, diffusion barrier components and shell-associated factors have all been localized using immunogold methods in green algae, diatoms and chlorarachniophytes \cite{McKay1991,Moromizato2024,Shimakawa2024,Yamano2010}. 
These experiments have established that pyrenoids possess complex spatial organization rather than behaving as homogeneous condensates. However, since chemical fixation TEM artefacts can distort hydrated biological assemblies, particularly condensate-like structures, more advanced electron microscopy methodologies have increasingly been applied to preserve native pyrenoid architecture. Large-scale fluorescent tagging pipelines have identified pyrenoid-associated proteins across multiple species \cite{mackinder_spatial_2017,Moromizato2024,Nam2024}. This extensive pyrenoid biomarker library has enabled a range of high-precision fluorescence microscopy experiments which demonstrate  highly organized spatial patterning within the pyrenoid and dynamic reorganization during environmental adaptation and cell division. Such microscopy hinted at a liquid-like state of the pyrenoid, confirmed through fluorescence recovery after photobleaching investigations \cite{freeman_rosenzweig_eukaryotic_2017}.

Although the molecular mechanisms of pyrenoid condensation may differ across species, including hornworts \cite{Robison2025}, pyrenoid assembly in the model alga \textit{Chlamydomonas reinhardtii} is driven primarily by multivalent interactions between Rubisco and the intrinsically disordered linker protein EPYC1 \cite{mackinder_coexistent_2016,he_structural_2020}. At the molecular scale, these interactions provide a natural setting for investigating the `sticker-and-spacer' framework of biomolecular condensation \cite{choi_physical_2020,freeman_rosenzweig_eukaryotic_2017,grandpre_impact_2023,payne-dwyer_predicting_2024,Kumar2026}. At a more coarse-grained level, recent work has shown how enzymatic regulation of effective EPYC1–Rubisco interactions can control pyrenoid condensate size, using a mean-field description of the underlying physics \cite{he_key1_2026}. Together, these developments position the pyrenoid as a promising model system for investigating the biophysical principles underlying cellular control of biomolecular condensates \cite{Brangwynne2015}.

The combination of a well-defined physiological function, tractable genetics, known molecular interactions and quantitative experimental observables makes the pyrenoid unusually amenable to comparison between theory and experiment.
Thus, whereas the biology and the physical `carbon concentration mechanism' of the pyrenoid has been thoroughly reviewed in several recent articles \cite{barrett_pyrenoids_2021,HeReview23,Catherall2025}, here we use the pyrenoid as a model system through which to discuss broader biophysical principles governing functional compartmentalization and biological self-organization. We first consider how transport, reaction and compartment architecture influence the performance of compartments, emphasizing reaction--diffusion descriptions and emerging questions concerning non-equilibrium operation. We then discuss how concepts from statistical mechanics and soft matter physics have been used to understand pyrenoid assembly, focusing on mean-field descriptions, multivalent interactions and (still hypothetical) scaffold-assisted condensation. Finally, we examine how cutting-edge experimental advances have enabled quantitative interrogation of both pyrenoid architectures and dynamic interactions, and discuss the broader lessons that pyrenoids may provide for understanding compartmentalization across biology.

\section{Biophysics of Catalytic Compartmentalization}

\subsection{The Pyrenoid as a Catalytic Microreactor}

The pyrenoid can be viewed as a catalytic microreactor\cite{greber_bacterial_2020} whose performance is determined not only by Rubisco kinetics but also by the delivery of inorganic carbon, removal of reaction products and suppression of competing pathways.
Their function therefore emerges from the interplay between transport, reaction and compartment architecture rather than enzyme activity alone \cite{halatek_rethinking_2018}.

The pyrenoid addresses the key challenge of Rubisco exhibiting relatively modest catalytic rates and additionally catalyzing a competing oxygenation reaction that initiates photorespiration \cite{andersson_backlund_2008,von_caemmerer_kinetics_1994,tcherkez_despite_2006}. Consequently, carbon fixation depends strongly on the local concentration of CO$_2$ surrounding the enzyme.

Maintaining elevated CO$_2$ concentrations is challenging because inorganic carbon continuously diffuses throughout the chloroplast and may be lost to the external environment. Many photosynthetic organisms therefore employ carbon concentrating mechanisms (CCMs), which actively accumulate inorganic carbon and elevate CO$_2$ concentrations near Rubisco \cite{barrett_pyrenoids_2021,HeReview23,Villarreal12}. Although their molecular implementations differ substantially across taxa, CCMs generally combine active transport, enzymatic interconversion of bicarbonate and CO$_2$, and spatial compartmentalization of carbon fixation.

In many algae and hornworts, this function is achieved by concentrating Rubisco within the pyrenoid, a dense proteinaceous compartment situated within the chloroplast \cite{barrett_pyrenoids_2021,Villarreal12}. The pyrenoid therefore serves as the central catalytic compartment of numerous CCMs, coupling carbon transport, enzymatic interconversion and carbon fixation within a single physical structure. This perspective forms the basis for the quantitative descriptions discussed below.

\subsection{Transport, Reaction and Leakage}
The importance of balancing carbon supply, fixation and leakage has long been recognised within the CCM literature \cite{Spalding2008,Moroney2007,Hanson2003}.
Early conceptual descriptions therefore viewed the pyrenoid primarily as a compartment that enhances carbon fixation by increasing carbon supply to Rubisco while suppressing diffusive losses.
What has changed in recent years is not this conceptual picture, but the level of molecular and structural detail available for quantitative analysis. Genetic, biochemical and structural studies have identified many of the transporters, enzymes and permeability barriers that contribute to pyrenoid function \cite{Yamano2010,wang_acclimation_2014,mackinder_coexistent_2016,Jin2016}. 
Collectively, these studies have produced a sufficiently detailed molecular inventory that transport, enzymatic interconversion and permeability barriers can now be incorporated explicitly into reaction--diffusion models \cite{fei_modelling_2022,kaste_reaction-diffusion_2024,fei_pyrenoid-based_2025}.

A particularly important advance of recent work by Fei and coworkers has been the integration of experimentally determined molecular components into quantitative reaction--diffusion descriptions of pyrenoid function \cite{fei_modelling_2022,fei_pyrenoid-based_2025}. These models explicitly incorporate inorganic carbon transport, carbonic anhydrase-mediated interconversion, Rubisco kinetics and permeability barriers, thereby providing a framework through which the functional consequences of molecular organization can be evaluated quantitatively. Such approaches have made it possible to identify conditions under which fixation becomes transport-limited, to quantify leakage pathways and to assess how pyrenoid architecture influences overall CCM performance.

These developments mark a transition from conceptual descriptions of catalytic compartmentalization to predictive mechanistic models of pyrenoid function. This naturally raises a further question: once sufficiently detailed mechanistic descriptions become available, can they themselves be reduced to simpler physical descriptions that capture the essential dynamics of catalytic compartment function? One way to address this question is to examine how reaction--diffusion models respond to controlled perturbations. Since algae experience continual fluctuations in both light availability and inorganic carbon supply \cite{Goldstein15}, dynamically varying environments provide a natural setting in which to identify effective parameters and reduced constitutive descriptions.

\subsection{Environmental Fluctuations and Nonlinear Dynamics}

Although Fei's reaction--diffusion model was developed to describe the molecular processes and cellular architecture of \textit{C. reinhardtii}, its response to controlled perturbations can nevertheless reveal coarse-grained properties that are expected to apply more broadly. This strategy of extracting effective behaviour from detailed mechanistic models is widely employed across physics \cite{Larson1999,McLeish2002,ChaikinLubensky}.
For example, oscillatory shear has long been used to probe viscoelastic materials, electrical circuits are characterized through their frequency response, and photosynthetic light reactions have been interrogated using fluctuating-light protocols to identify characteristic timescales and regulatory mechanisms \cite{Acevedo-Siaca2025-sb,AbuGhosh2016,Schulze2017Flashing}. More generally, dynamic perturbations provide a route to infer effective parameters that may not be accessible from steady state measurements alone. From an engineering perspective, perturbations may even be beneficial to design reactors whose average performance exceeds that of reactors operated in steady state  \cite{SilvestonHudgins2013}.

In this spirit, algae in photobioreactors have previously been subjected to flashing light sources, enabling new understanding into the mechanisms by which metabolites are buffered in the cell \cite{Acevedo-Siaca2025-sb,AbuGhosh2016,Schulze2017Flashing}.
Using the developments of reaction-diffusion models, we may now ask how pyrenoid-associated processes respond to environmental fluctuations 
\cite{fei_modelling_2022,kaste_reaction-diffusion_2024,fei_pyrenoid-based_2025}, and if the full reaction--diffusion model can be coarse-grained to a minimal constitutive equation 
\begin{equation}
\frac{\mathrm{d}c}{\mathrm{d}t}
=
R_\mathrm{supply}(t)
-
R_\mathrm{fixation}(c)
-
R_\mathrm{leakage}(c),
\end{equation}
that encompasses the general need for balance between supply, fixation, and leakage in pyrenoids of any species \cite{Moroney2007,Spalding2008}.

In this equation, \(c\) represents a concentration of available inorganic carbon (which need not to be the \(\mathrm{CO_2}\) concentration itself, but may include a fraction of buffered \(\mathrm{HCO_3}^-\) that could be rapidly converted). 
Admittedly, the supply rate \(R_\mathrm{supply}(t)\) would be difficult to precisely control experimentally, but fundamentally for any small amplitude sinusoidal fluctuation there will be a time-averaged fixation rate that approximates to 
\begin{equation}\langle R_\mathrm{fixation}\rangle=\alpha-\frac{\beta}{1+\omega^2\tau^2},\label{eq:toymodel}
\end{equation}  
where the coefficients \(\alpha\) and \(\beta\), and the relaxation time \(\tau\), are effective parameters determined by the local transport and reaction kinetics near steady state. 
This approach is analogous to frequency-response analyses used throughout physics to extract constitutive parameters and characteristic relaxation times from complex systems.
Here, \(\alpha-\beta\) and \(\alpha\) correspond to the low- and high-frequency asymptotes, respectively, while \(\tau\) defines an effective relaxation timescale. Importantly, the low-frequency asymptote \(\alpha-\beta\) need not equal the fixation rate evaluated at the average concentration. This reflects the nonlinear dependence of fixation and leakage on carbon concentration and implies that slow oscillations can alter the average performance even when the mean supply remains unchanged.

\begin{figure}[ht!]
\centering
\hphantom{0} \hfill
\begin{minipage}{0.4\textwidth}
\centering
\llap{\textbf{a}\hspace{2pt}}\vspace{-2em}
\includegraphics[width=\textwidth]{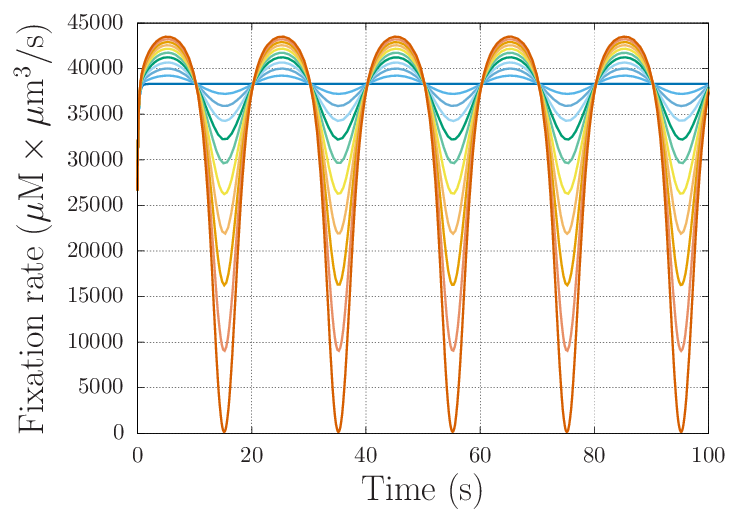}
\end{minipage}
\hspace{1cm}
\begin{minipage}{0.4\textwidth}
\centering
\llap{\textbf{b}\hspace{2pt}}\vspace{-2em}
\includegraphics[width=\textwidth]{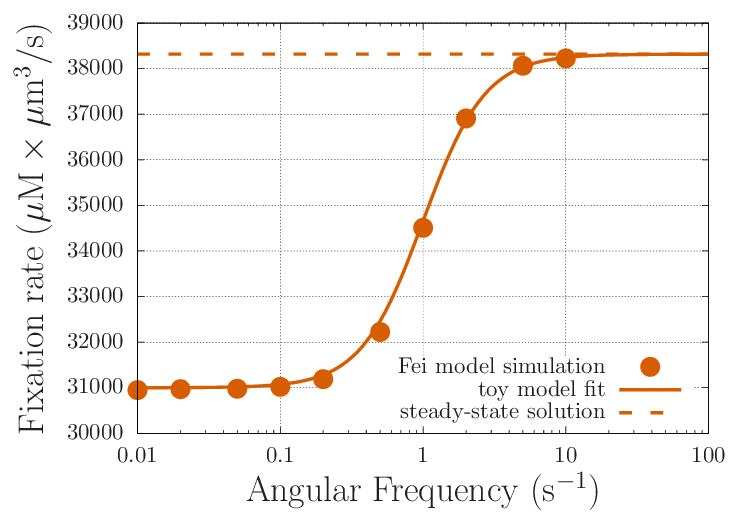}
\end{minipage}
\hfill \hphantom{0}
\caption{ 
Fixation rate in response to oscillations in supply as predicted by a time-dependent extension of the reaction--diffusion model of Fei \textit{et al.}.
(a) Transient fixation rate for fixed
\(\omega=0.1\,\mathrm{s^{-1}}\) and amplitudes increasing from
\(A=0\) to \(A=1\). For small amplitudes the response is approximately
sinusoidal, whereas for larger amplitudes the peaks become increasingly
suppressed owing to saturation of uptake and/or fixation processes. 
(b) Time-averaged fixation rate for fixed \(A=1\) and varying
\(\omega\). The symbols were calculated using the full reaction-diffusion model; the solid curve is the toy model in Eq.~\ref{eq:toymodel} fitted to the symbols; the dashed curve corresponds with the steady-state
value (i.e., for amplitude \(A=0\)).
}
\label{fig:oscillations}
\end{figure}

Recent advances in quantitative reaction-diffusion modelling now make it possible to explore dynamical questions that extend beyond steady-state carbon fixation.
As an illustration, one may consider  extending the steady-state framework of Fei et al. to time-dependent perturbations.
As a representative example, oscillatory fluctuations in cytosolic inoragnic carbon supply can be imposed according to 
\(c=c_0(1+A\sin(\omega t))\), where \(c_0\) is the average concentration
(\(10\,\mu\mathrm{M}\) for CO$_2$ and \(100\,\mu\mathrm{M}\) for
H$_2$CO$_3$/HCO$_3^-$), \(A\) is the relative amplitude and \(\omega\)
is the angular frequency. 
Solving the corresponding time-dependent extension of the reaction--diffusion model, including starch barriers and passive (rather than active) transport through chloroplast envelope \(\mathrm{HCO}_3^-\) transporter LCIA, enables the transient fixation dynamics to be analysed (Figure~\ref{fig:oscillations}\textbf{a}). The resulting frequency-dependent average fixation rates (Figure~\ref{fig:oscillations}\textbf{b}) are well described by the simple constitutive expression in Eq.~\ref{eq:toymodel}, illustrating how detailed mechanistic models can be coarse-grained into effective dynamical descriptions.

The agreement between the full reaction–diffusion model and the reduced constitutive description is encouraging, but should be interpreted with caution, as it likely points at a decoupling of biochemical processes in the model (e.g., due to the choice of rapid conversion rates), rather than the emergence of a  single effective timescale.
Nevertheless, this analysis illustrates how recent advances have put pyrenoid research in a place where the theoretical toolkits from soft-matter science, in particular coarse-graining techniques, can be applied to test hypotheses on organelle function.

\section{Biophysics of Compartment Self-Organization}
\subsection{From Molecular Interactions to Compartments}

Biomolecular condensates are frequently interpreted using two complementary theoretical frameworks. Mean-field descriptions treat condensation as a thermodynamic phenomenon governed by phase coexistence, criticality and nucleation, whereas sticker-and-spacer models seek to explain how such behaviour emerges from specific molecular interactions encoded by macromolecular architecture \cite{choi_physical_2020}. Together, these approaches provide a conceptual bridge between molecular interactions and mesoscale compartment formation.

An important challenge is to connect these levels of description. Mean-field theories provide experimentally testable phase diagrams and a compact thermodynamic description of condensation, whereas molecular descriptions seek to explain how such thermodynamic behaviour arises from specific interactions between proteins and nucleic acids. Establishing quantitative links between molecular architecture and phase behaviour remains an active area of research.

Pyrenoids provide a particularly attractive system in which to address these questions. Unlike many condensates, whose composition and interactions remain incompletely characterized, pyrenoid assembly is dominated by a relatively small number of experimentally tractable interactions. As a result, pyrenoids offer a rare opportunity to investigate both the molecular mechanisms that encode condensation and the thermodynamic behaviour that emerges from them within the same biological system.

\subsection{What Do Pyrenoids Tell Us About Molecular Grammar?}

While phase separation was hypothesised as a physical basis for intracellular microcompartmentation in 1995 by Walter \& Brooks \cite{walter_phase_1995}, the modern field of biomolecular condensates remains to be challenged by questions on the  dynamic regulation of phase separation and the role of molecular grammar in this process; i.e., what are the sequence-encoded rules that dictate how the amino acid composition and patterning of proteins drive biomolecular condensation  \cite{mittag_conceptual_2022,lyon2021physics,alberti_biomolecular_2021}. A distinctive feature of the pyrenoid is that both the molecular interactions responsible for assembly and the resulting thermodynamic behaviour can be investigated quantitatively within the same biological system. Long before pyrenoids became widely discussed as biomolecular condensates, genetic, biochemical and structural studies had already identified specific interactions between Rubisco and the linker protein EPYC1 as central determinants of pyrenoid organization \cite{mackinder_coexistent_2016}.

\begin{figure}[ht!]
\centering
\hphantom{0} \hfill
\begin{minipage}{0.17\textwidth}
\centering
\llap{\textbf{a}\hspace{2pt}}\vspace{-2em}
\includegraphics[width=\textwidth, trim={40cm 36cm 40cm 36cm}, clip]{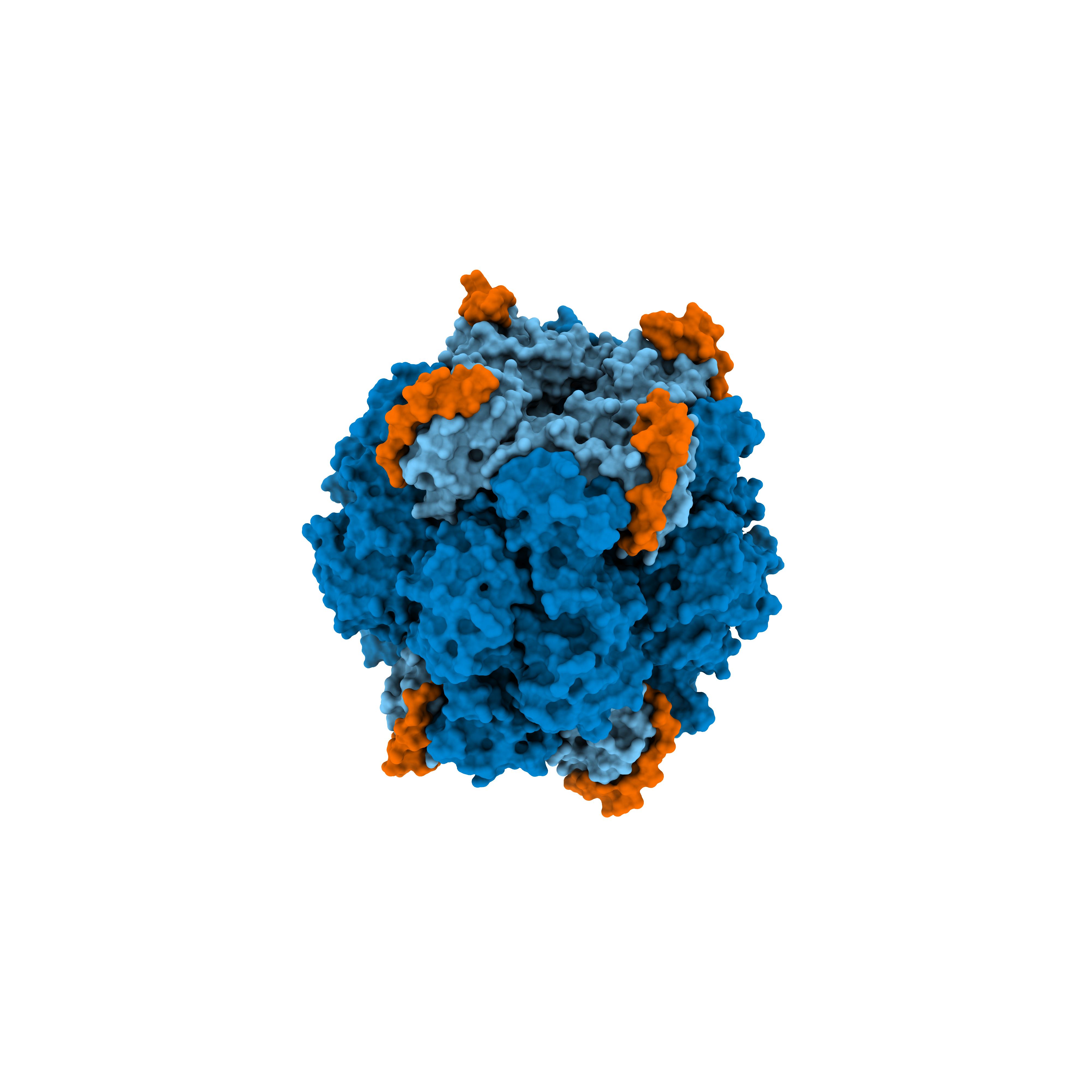}\\
\end{minipage}
\hspace{1cm}
\begin{minipage}{0.6\textwidth}
\centering
\llap{\textbf{b}\hspace{2pt}}\vspace{-2em}
\includegraphics[width=\textwidth]{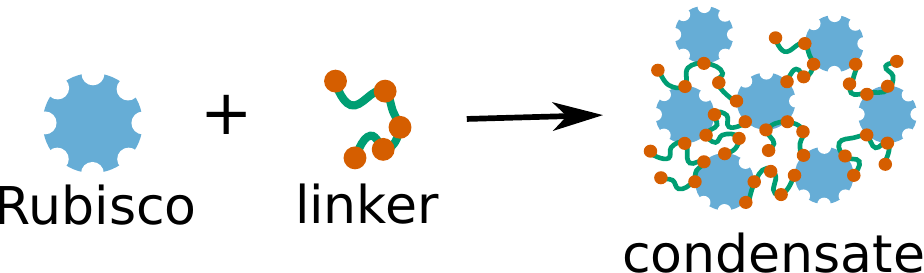}
\end{minipage}
\hfill \hphantom{0}
\caption{ (a) Rubisco protein structure (large and small subunits in dark and light blue, respectively) with bound EPYC1 peptides in orange (This cryo-EM structure was obtained from the Protein Data Bank (PDB ID: 7JFO) \cite{he_structural_2020}). (b) Schematic sticker-and-spacer representation of Rubisco:linker system.}
\label{fig:Rubisco}
\end{figure}

The discovery that pyrenoid assembly is governed by a comparatively small number of multivalent interactions created an opportunity to apply concepts from polymer physics and colloid science to a biological compartment with a well-defined physiological function \cite{choi_physical_2020}. 
In particular, the sticker-and-spacer framework provides a useful language for describing these interactions \cite{choi_physical_2020}. Within this picture, intermolecular binding is mediated by a limited number of interaction motifs (`stickers') connected by flexible regions (`spacers'), see Figure~\ref{fig:Rubisco}. Similar ideas have long been used in the study of associating polymers and patchy colloids, where macroscopic behaviour is governed by the number, strength and geometry of intermolecular bonds \cite{RUBINSTEIN199983,RubinsteinColby,Glotzer2007Anisotropy,Bianchi2011PatchyReview}. In biological systems, folded protein domains may act as discrete interaction sites analogous to patchy particles, while intrinsically disordered regions frequently serve as polymeric spacers whose flexibility modulates network formation \cite{choi_physical_2020,HolehouseKragelund2023}.

The sticker-and-spacer framework was first applied to the pyrenoid by Mackinder \textit{et al.} \cite{mackinder_coexistent_2016}, who proposed that EPYC1 acts as a multivalent linker that crosslinks Rubisco molecules into a dynamic network. This interpretation subsequently suggested a physical explanation for the coexistence of liquid-like behaviour and molecular specificity: condensate assembly is driven by many weak, reversible interactions rather than a small number of high-affinity binding events\cite{freeman_rosenzweig_eukaryotic_2017}. Subsequent biochemical and structural studies confirmed this picture. Surface plasmon resonance measurements demonstrated reversible binding between EPYC1 helices and Rubisco \cite{he_structural_2020}, while structural studies identified the corresponding interaction sites. 

The experimentally determined interactions subsequently enabled the development of coarse-grained molecular descriptions. Freeman-Rosenzweig and co-workers used molecular simulations to demonstrate how reversible Rubisco--EPYC1 interactions can generate a liquid-like condensate \cite{freeman_rosenzweig_eukaryotic_2017}. These models were later extended by GrandPre \textit{et al.} to investigate how the spatial arrangement of Rubisco binding sites and the length of linker molecules influence condensate formation \cite{grandpre_impact_2023}. Related approaches have also provided a framework for studying interactions between the pyrenoid matrix and the thylakoid membrane system \cite{Yu2024}. Collectively, these studies provided a framework for relating molecular architecture to condensate behaviour.

A central question is whether condensate behaviour can be predicted directly from molecular interactions. Recent work has come remarkably close to this objective. 
Rather than simply testing an existing theoretical framework, Payne-Dwyer \emph{et al.} developed a statistical-mechanical description in parallel with quantitative biophysical experiments specifically designed to determine the molecular parameters required by the theory \cite{payne-dwyer_predicting_2024}.
Building on measurements of dilute-phase Rubisco--EPYC1 complexes \cite{He2023}, binding affinities, sticker valencies and specer flexibility were quantified experimentally and incorporated into a statistical-mechanical framework to predict the concentration at which Rubisco dimerisation occures. Because dimer formation represents the first step towards network assembly, this concentration serves as a proxy for the onset of phase separation.

\begin{figure}[ht!]
\centering
\begin{minipage}{0.5\textwidth}
\centering
\llap{\textbf{a}\hspace{0cm}}\vspace{-2em}
\includegraphics*[width=\textwidth]{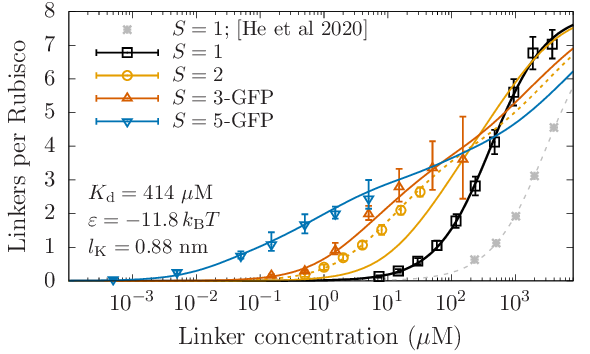}
\end{minipage}
\begin{minipage}{0.4\textwidth}
\centering
\llap{\textbf{b}\hspace{0cm}}\vspace{-2em}
\includegraphics[width=\textwidth]{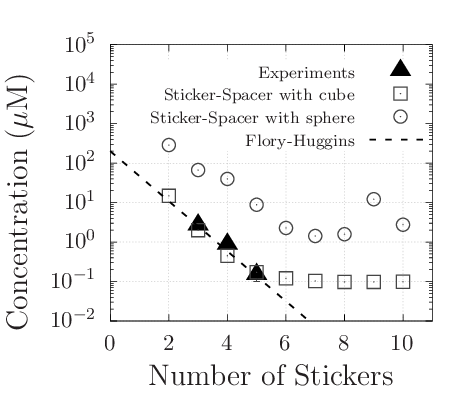}
\end{minipage}
\caption{\textbf{a}. Number of bound linkers per Rubisco monomer in the dilute phase, measured using SPR and Slimfield microscopy \cite{Plank2009Slimfield} for various numbers of stickers per linker. The curves represent statistical theory fitted to the experiment using the dissociation constant, binding constant \(K_\mathrm{d}\) and the Kuhn length \(l_\mathrm{k}\) as a measure for the spacer flexibility (reproduced from Ref.\citenum{payne-dwyer_predicting_2024}). \textbf{b}. The parameters determined in \textbf{a} were used to predict the minimum binodal concentration using two different geometric representations of Rubisco (sphere and cube). (reproduced from Ref.\citenum{Kumar2026}). The dashed curve was calculated using the phase diagrams in Figure 5; see section~\ref{sec:Thermo}.
}
\label{fig:titration}
\end{figure}

The resulting predictions were in quantitative agreement with experimental observations across a series of engineered linker constructs (Figure~\ref{fig:titration}). More broadly, this work illustrates how statistical-mechanical theory can be used not only to interpret experiments but also to guide the design of quantitative measurements that determine the molecular parameters governing condensate assembly. It therefore represents an important step towards predictive descriptions in which molecular properties measured in the dilute phase can be used to forecast collective assembly behaviour.

At the same time, subsequent work clarified both the predictive power and the limitations of current molecular descriptions.
Kumar \textit{et al.} showed that the predicted phase-separation concentrations depend sensitively on assumptions regarding Rubisco geometry and interaction topology \cite{Kumar2026}. In particular, application of the framework to \textit{Chlorella} substantially overestimated the experimentally observed concentrations, suggesting that cooperative assembly processes or additional interactions may contribute to condensate formation.

More generally, molecular simulations and statistical-mechanical models typically characterize assembly through proxies such as dimerization, cluster formation or network connectivity. By contrast, the full phase behaviour of the system, including coexistence boundaries, critical points and assembly pathways, remains difficult to obtain from molecular descriptions alone. Thus, while molecular coarse-graining successfully connects specific interactions to condensate assembly, it does not yet provide a complete description of phase behaviour.

More generally, molecular simulations and statistical-mechanical models are primarily concerned with the emergence of molecular connectivity, describing assembly through quantities such as dimerization, cluster formation or network topology. These observables provide valuable insight into the microscopic origins of condensation but do not by themselves determine macroscopic phase behaviour, including coexistence boundaries, critical points or nucleation pathways. Addressing these collective properties requires an additional level of coarse-graining in which molecular complexity is represented through effective thermodynamic interactions.


\subsection{What Do Pyrenoids Tell Us About Thermodynamics?}\label{sec:Thermo}

The observation that the pyrenoid is liquid-like \cite{freeman_rosenzweig_eukaryotic_2017} has established it as an important model system within the broader family of biomolecular condensates \cite{Brangwynne2015,shin_liquid_2017}. This has created opportunities to apply concepts from equilibrium thermodynamics and soft matter physics to pyrenoid assembly \cite{choi_physical_2020,RubinsteinColby,deGennes}. Within this framework, specific molecular interactions are coarse-grained into a small number of effective parameters that determine collective behaviour. Consequently, the relevant observables are no longer individual binding events, but phase boundaries, critical points, tie lines, coexisting phase compositions and condensate dynamics.

Recent work has demonstrated that pyrenoid behaviour can indeed be understood and quantitatively predicted using such effective thermodynamic descriptions. Indeed, He and coworkers  showed that phosphorylation of EPYC1 leads to pyrenoid dissolution in vivo \cite{he_key1_2026}, and using a mean-field thermodynamic framework, the observed changes in condensate size could be interpreted through phosphorylation-induced changes in an effective interaction strength, rather than through a detailed molecular description of every sticker, spacer and post-translational modification. More broadly, these results establish the pyrenoid as a tractable model system in which the physical principles governing condensate regulation can be investigated quantitatively.

\subsubsection{From molecular sticker-and-spacer models to thermodynamics}~
Mean-field thermodynamic models also provide a useful conceptual framework for asking how changes in molecular properties influence condensate assembly. As an illustration, a minimal three-component Flory-Huggins description provides a useful conceptual framework for relating molecular design parameters to phase behaviour while highlighting the strengths and limitations of coarse-grained descriptions.
In this framework, the free energy is written as \cite{Schaefer16,he_key1_2026},
\[
f=\frac{\phi_\mathrm{rub}\ln\phi_\mathrm{rub}}{N_\mathrm{rub}}+\frac{\phi_\mathrm{link}\ln\phi_\mathrm{link}}{N_\mathrm{link}}+\frac{\phi_\mathrm{w}\ln\phi_\mathrm{w}}{N_\mathrm{w}}
+\phi_\mathrm{rub}\phi_\mathrm{link}\chi,
\]
where \(\phi_\mathrm{rub}\), \(\phi_\mathrm{link}\) and \(\phi_\mathrm{w}\) denote the volume fractions of Rubisco, linker and water, respectively. Following He \textit{et al.}, the  interaction parameter \(\chi\) represents an effective Rubisco-linker interaction strength.
To illustrate how molecular design parameters influence phase behaviour within such a coarse-grained framework, the linker size (\(N_\mathrm{link}\)) can be varied to mimic the engineered sticker constructs examined experimentally.

\begin{figure}[ht!]
\centering
\begin{minipage}{0.45\textwidth}
\centering
\llap{\textbf{a}\hspace{0cm}}\vspace{-2em}
\includegraphics*[width=\textwidth]{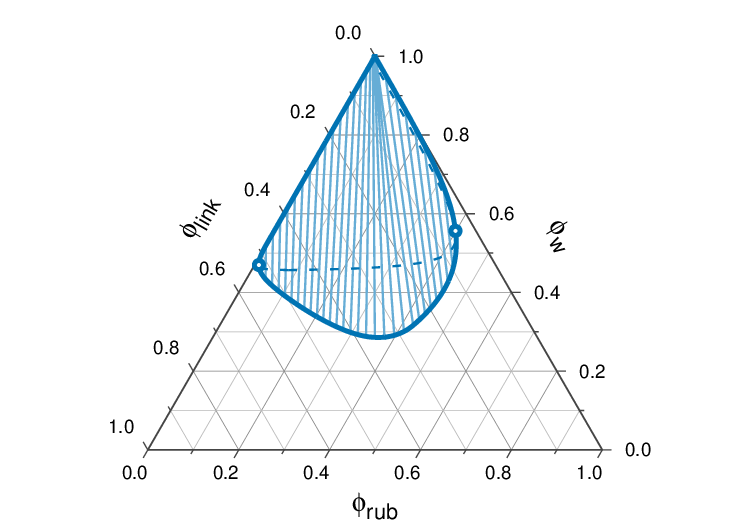}
\end{minipage}
\begin{minipage}{0.45\textwidth}
\centering
\llap{\textbf{b}\hspace{0cm}}\vspace{-2em}
\includegraphics[width=\textwidth]{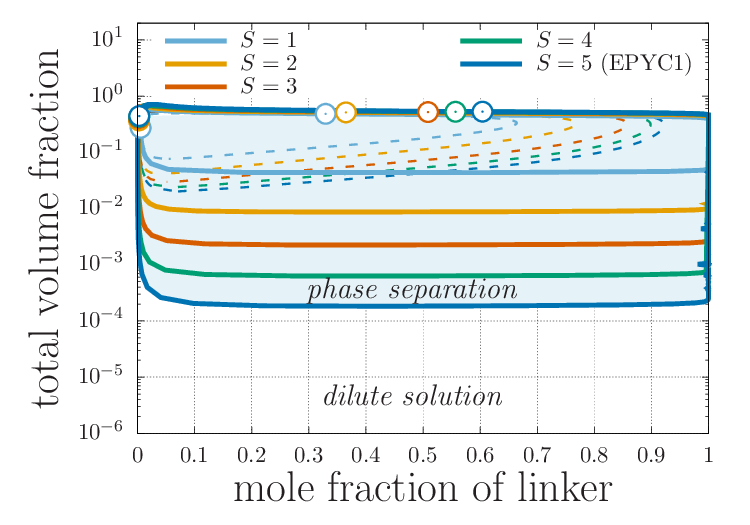}
\end{minipage}
\caption{Flory-Huggins phase diagram for the LLPS of Rubisco:linker:water systems with \(\chi=-4.5\), \(N_\mathrm{rub}=348\) and \(N_\mathrm{link}=2S\). In (a) \(S=5\) represents EPYC1. In this ternary diagram, the solid and dashed curves are the binodal and spinodal, respectively. The straight lines are the tie lines, and the circles represent the  critical points. 
(b) Same phase diagram as (a) but including sticker numbers \(S=1-5\); to focus on dilute regime, the total volume fraction (\(\phi_\mathrm{link}+\phi_\mathrm{rub}=1-\phi_\mathrm{w}\)) is plotted against the mole fraction of linker (\(\phi_\mathrm{link}/(\phi_\mathrm{link}+\phi_\mathrm{rub})\)). For increasing sticker number, the spinodal region (enclosed by dashed curves) is relatively unaffected,  while the binodal exponentially shifts to lower concentrations; this shift is indicated by the dashed line in Figure 4 b in the previous section. }
\label{fig:tern}
\end{figure}

The resulting phase diagrams are shown in Fig.~\ref{fig:tern}. 
Despite its simplicity, this minimal description reproduces several experimentally observed trends.
In particular, the predicted dependence of the lower binodal concentration on sticker number agrees quantitatively with microscopy observations that were previously interpreted using sticker-and-spacer theory \cite{payne-dwyer_predicting_2024,Kumar2026}. Thus, despite the considerable molecular complexity of the pyrenoid matrix, key features of assembly appear to be describable in terms of only a small number of effective parameters, namely interaction strength and molecular size, albeit with the caveat that caution is needed in generalizing the findings from many of the in vitro measurements to the pyrenoid in vivo.

Like all coarse-grained descriptions, however, this simplification is achieved at the cost of molecular detail. The mean-field phase diagrams predict substantial tunability of dense-phase composition with stoichiometry and concentration, whereas experimentally the Rubisco concentration within pyrenoids appears to be remarkably robust \cite{Holdsworth68,Kowallik69,Bertagnolli70,Engel15,Griffiths1970,meyer2012pyrenoid,freeman_rosenzweig_eukaryotic_2017,he_structural_2020}. Likewise, the theory predicts that single-sticker linkers may induce phase separation, whereas such constructs are experimentally used precisely because they do not phase separate \cite{he_structural_2020}. These discrepancies arise because specific Rubisco--EPYC1 interactions are replaced by a single effective interaction parameter. Consequently, the theory captures generic thermodynamic trends while failing to describe aspects of assembly that depend on interaction specificity.

Despite these limitations, the mean-field framework successfully captures several experimentally observed features of pyrenoid assembly and regulation. In particular, it explains how condensate behaviour can be tuned through changes in effective interaction strength \cite{he_key1_2026} and correctly predicts the reduction in phase-separation concentration with increasing sticker number. Together, these results illustrate how coarse-grained thermodynamic descriptions can connect molecular perturbations, condensate regulation and assembly pathways through a small number of effective parameters.

\subsubsection{Nucleation and hypothesis of scaffold-assisted self-assembly}~
An additional insight emerging from the phase diagrams is that physiologically relevant pyrenoids (volume fractions \(\ll 0.1)\) lie far from both the critical point and the spinodal region. This suggests that phase separation proceeds through nucleation and growth rather than spontaneous spinodal decomposition. In such a regime, condensate formation requires fluctuations to overcome a free-energy barrier. While this may occur through thermal fluctuations alone, large activation barriers may in practice favour nucleation on pre-existing cellular structures.

This distinction is particularly relevant in living cells because nucleated assembly provides a natural route for spatial control. Whereas spinodal decomposition occurs spontaneously throughout an unstable region, nucleated assembly can be localized through structures that lower the activation barrier. In the pyrenoid, the close association between the Rubisco matrix and a network of thylakoid tubules has long suggested a coupling between condensate assembly and chloroplast architecture \cite{Engel15,mackinder_spatial_2017}. 
Indeed, several membrane-associated Rubisco-binding proteins (RBMPs) were identified on these tubules and shown to interact with pyrenoid components \cite{he_structural_2020,Hennacy2024,Yu2024}, and which leads to the suggestion that thylakoid-associated structures may contribute to pyrenoid assembly.

The thermodynamic arguments presented above provide an independent motivation for considering such mechanisms. If physiologically relevant pyrenoids indeed form outside the spinodal region, condensate formation may require nucleation events that are facilitated by pre-existing cellular structures. However, the molecular basis of such nucleation remains unresolved. In particular, recent work demonstrated that BST4 (RBMP1), previously proposed to participate in pyrenoid organization, is not required for pyrenoid tubule formation and instead appears to function primarily in chloroplast ion homeostasis and acclimation to fluctuating light \cite{Adler24}. Thus, while biological and thermodynamic considerations both point towards a possible role for cellular architecture in pyrenoid assembly, direct evidence for scaffold-assisted nucleation remains limited.

More broadly, the examples discussed above illustrate how thermodynamic coarse-graining can connect molecular interactions, condensate regulation and assembly pathways through a small number of effective parameters. In this respect, the pyrenoid has emerged as an unusually tractable system in which the biophysics of biomolecular condensates can be investigated quantitatively. Whether similar thermodynamic descriptions can ultimately explain how cells control the location and timing of condensate assembly remains an important open question.

\section{Innovations in Biophysical Experimentation}
Innovations in experimental studies are important since they have the capacity to  reveal new biology. However, just as importantly, developments in experimental advances can increasingly enable quantitative physical models to be parameterized and tested.  Recent transformative experimental biophysics advances relating to the pyrenoid  have focused on two broad themes: developments in our understanding of its architecture and of the dynamic interactions between the Rubisco and linkers.

\subsection{Advances in Understanding Pyrenoid Architectures}
Quick-freeze deep-etch electron microscopy preserves more native structural organization without chemical fixation \cite{mackinder_coexistent_2016}. By rapidly vitrifying cellular material before fracture and metal shadowing, this approach has revealed pyrenoid surface topology and their mesoscale organization in greater physiological detail and showed that the pyrenoid matrix behaves as a highly interconnected network of macromolecular interactions rather than as a crystalline aggregate.
More recently, cryo-electron tomography (cryo-ET) has revolutionized understanding of pyrenoid architecture by enabling nanoscale 3D visualization of vitrified cellular material in near-native states, bridging the gap between classical structural biology and mesoscale cellular organization.  \cite{Elad2025,Engel15,Nam2024,Shimakawa2024}. Cryo-ET studies have resolved pyrenoid tubule systems, matrix organization and peripheral diffusion barriers with unprecedented detail. In diatoms, cryo-ET revealed highly ordered proteinaceous shell structures surrounding pyrenoids \cite{Shimakawa2024}, while studies in green algae demonstrated intricate membrane tubule geometries embedded within the condensate matrix.
The application of subtomogram averaging has enhanced the structural power of cryo-ET by exploiting repeated structural motifs within tomograms to improve signal-to-noise ratios and reconstruct protein assemblies at much higher resolution. Recent work is resolving pyrenoid-associated proteins directly within their native intracellular context using these methods \cite{Shimakawa2024}.  Serial block-face scanning electron microscopy has complemented these nanoscale methods by enabling volumetric 3D reconstructions over larger cellular scales \cite{Atkinson2024,Hennacy2024}. These approaches allow pyrenoid architecture to be contextualized within entire chloroplasts and cells, revealing spatial relationships between pyrenoids, thylakoid membranes and other organelles. This approach has been particularly useful for characterizing mutant phenotypes in CCM-deficient strains and examining how pyrenoid morphology responds to environmental conditions.

\subsection{Developments in Understanding Rubisco-Linker Interactions}
Understanding the molecular interactions underlying pyrenoid assembly has required increasingly sophisticated biophysical methodologies. Rubisco-linker interactions present substantial experimental challenges because full-length linker proteins exhibit high valency and readily phase separate at physiological concentrations.
Consequently, many studies have focused on isolated linker fragments representing individual binding motifs. SPR \cite{Barrett2024,he_structural_2020}, bio-layer interferometry \cite{Oltrogge2020} and microscale thermophoresis have all been applied to characterize these weak interactions quantitatively. These experiments revealed that individual Rubisco-linker interactions are typically very low affinity, often in the high micromolar to millimolar range. Such weak interactions are nevertheless highly effective at driving condensation when combined through multivalency.
Cryo-electron microscopy approaches have also been used to determine structural details of Rubisco-linker interactions. Because Rubisco is large, highly symmetric and structurally stable, it is particularly amenable to cryo-EM reconstruction. However, the low affinity of linker interactions complicates structural characterization. Various strategies have therefore been employed, including the use of multivalent fragments, high peptide concentrations and engineered fusion constructs \cite{Barrett2024,he_structural_2020,Oh2023,Oltrogge2020}.
Recent improvements in detection sensitivity have enabled study of full-length linker proteins under dilute conditions below the phase separation threshold. Fluorescence correlation spectroscopy (FCS) \cite{He2023} and single-molecule binding assays \cite{payne-dwyer_predicting_2024} have provided new quantitative insight into interaction stoichiometry and kinetics. Nevertheless, direct experimental characterization of Rubisco-linker interactions within the condensed phase itself remains limited.
Future biophysical approaches will likely increasingly exploit spectroscopic techniques already widely applied in biomolecular condensate research. Integrating such experimental data with computational modeling approaches that explicitly incorporate multivalent cross-linking and phase behavior will be critical for developing predictive physical models of pyrenoid organization \cite{grandpre_impact_2023,payne-dwyer_predicting_2024}.

\begin{figure}[ht!]
\centering
\includegraphics*[width=\textwidth]{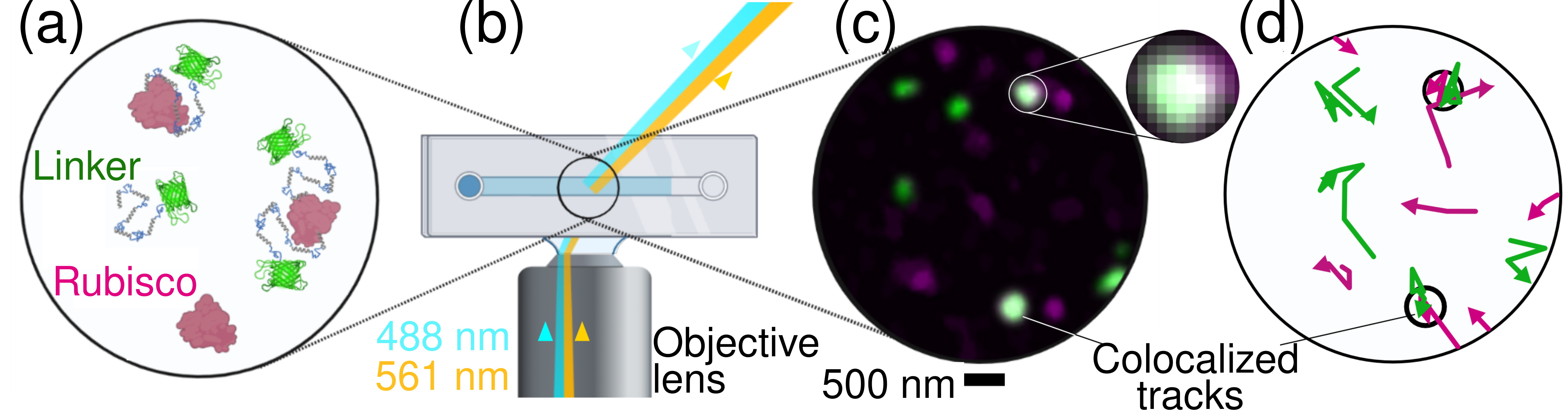}
\caption{Quantitative binding of linker and Rubisco using Slimfield. (a),(b) Rubisco-Atto594 is equilibrated with linker at mutual
concentrations insufficient for phase separation, and introduced to
a simple microscope chamber. (c) Slimfield reveals how assemblies of Rubisco (magenta, max projection) and/or linker GFP
(green) adsorb transiently and nonspecifically to the cover glass.
(d) Rapid molecular motion is reconstructed into tracks and unique
colocalizations. Reproduction from Ref.~\citenum{payne-dwyer_predicting_2024}}
\label{fig:slimfield}
\end{figure}


\section{Concluding Remarks}

\subsection{Emerging Frontiers in Pyrenoid Biophysics} 
A key challenge to understanding the biophysics of real biomolecular condensates in situ, namely inside a functioning, living cell, lies in their ubiquity across multiple species in all domains of life, and being implicated across several important cellular processes. This pervasive quality is manifest in biological complexity and significant apparent heterogeneity in regard to the molecular components involved
resulting in difficulties in comparing research findings across different studies. One strategy to mitigate this issue is for the condensate research community to agree on standardized protocols  and to converge on  standards for both reagents and model systems where possible, with the caveat that there may be unavoidable biochemical reasons which prohibit convergence.  The pyrenoid is emerging as an excellent model system; it has evolved across multiple organisms from different biological start points but has converged upon consensus biophysical features.  Across the  pyrenoids from different species there are broad ranges exhibited for each relevant biophysical parameter, such as  fluidity in ranging from a  liquid state to far more glassy and sometimes solid states; this breadth of parameter space is a useful feature to gaining understanding concerning how molecular scale interactions emerge into mesoscale material properties. And thirdly, pyrenoids exemplify well how the biophysical principles of biomolecular condensates in situ are governed not just by the condensate itself but also by its physicochemical environment, illustrated by the rich spatial patterning of the surrounding starch sheath and the associated temporal dynamics of \(\mathrm{CO_2}\) flux across it. There are several strong arguments for urging the condensate biophysics community to consider the benefits of more research both driven from experiment and theory focused around the pyrenoid as a model system not only for a biomolecular condensate in situ but more generally for studying the biophysics of spatiotemporal compartmentalization and organization in a living cell.

Collectively, the statistical physics, computational and experimental biophysics approaches we have discussed in this review have transformed the pyrenoid from a poorly understood chloroplast inclusion into one of the best-characterized biomolecular condensates. Importantly, pyrenoid research now occupies a unique position at the intersection of cell biology, structural biology, soft matter physics and biophysics. Increasingly, experimental efforts are moving beyond simple compositional descriptions towards quantitative understanding of emergent physical behavior. Key future questions include how pyrenoid material properties influence \(\mathrm{CO_2}\) fixation efficiency, how condensate dynamics are regulated in vivo, how phase behavior integrates with membrane organization and how pyrenoid architectures evolved convergently across diverse lineages. Addressing these challenges will require continued integration of advanced microscopy, quantitative proteomics, structural biology, spectroscopy, single-molecule methods, computational biophysics and pencil-and-paper statistical physics theory. The pyrenoid therefore represents not only a central system for understanding photosynthetic carbon fixation, but also an increasingly powerful model for studying the physical principles governing intracellular organization in living systems.

A recurring theme throughout this review is that increasing molecular and structural complexity does not necessarily require increasingly complex physical descriptions. Reaction--diffusion models suggest that catalytic performance can often be understood through the balance between a small number of competing fluxes and characteristic timescales, while statistical-mechanical descriptions indicate that condensate assembly may be governed by a limited set of molecular design parameters and thermodynamic constraints. Whether such reduced descriptions represent a special feature of pyrenoids or reflect more general principles governing biological compartments remains unclear. Nevertheless, the combination of a well-defined physiological function, extensive molecular characterization and growing theoretical framework makes the pyrenoid an unusually attractive system in which to investigate this possibility as the research community moves to the next stage of exploring more condensates in their natural cellular environments as well as synthetic condensates integrated into living cells.

\begin{summary}[SUMMARY POINTS]

\begin{enumerate}
\item  Unlike many other biomolecular condensates, the function of the pyrenoid is widely agreed to balance the competing processes of substrate supply, molecular leakage and catalytic conversion.

\item The molecular components of the pyrenoid have been sufficiently well characterised to develop reaction-reaction models that explain the macroscopic function of the pyrenoid in terms of both the architecture and the components.

\item Despite the molecular and architectural complexity, the mechanistic models specific to algae \emph{C. rheinhartdii}, may be generalised for compartmentalized systems through 
conventional coarse-graining techniques.

\item The molecular origins of the architecture appears to be controlled by a relatively small number of components using complementary molecular descriptions using the `sticker-and-spacer' framework and mean-field thermodynamics.

\item While active processes can dynamically control the thermodynamics, and thereby the size of the condensates, it remains unknown if active processes and/or scaffolds may affect the nucleation of condensates.


\item  New variants of single-molecule microscopy and associated analysis tools can reliably quantify time-resolved molecular interactions relevant to the pyrenoid.
\item Advances in tomography imaging methods, in particular application of cryo-ET, has recently enabled transformative new insights into the 3D  structure of condensates at exceptionally high spatial resolution.  
\end{enumerate}

\end{summary}


\section*{Acknowledgements}

ML and CS were supported by EPSRC grant EP/W024063/1; ML was additionally supported by EPSRC grant EP/Y000501/1.

We gratefully acknowledge Tom McLeish, whose early insights helped shape many of the biophysical perspectives presented in this review.

We thank James Barrett and Luke Mackinder (University of York) for their expert feedback on an earlier draft of the manuscript.


\bibliographystyle{unsrtnat}
\bibliography{references}

\end{document}